\documentclass[prd,aps,eqsecnum,tightenlines,nofootinbib,showpacs]{revtex4}
\input epsf
\usepackage{graphicx}

\newcommand{\ApJ}{Astrophys. J.}

\newcommand{\MNRAS}{Mon. Not. Roy. Astron. Soc.}
\newcommand{\AandA}{Astron. \& Astrop.}

\newcommand{\beq}{\begin{equation}}
\newcommand{\beqa}{\begin{eqnarray}}
\newcommand{\eeq}{\end{equation}}
\newcommand{\eeqa}{\end{eqnarray}}

\newcommand{\gsim}{\gtrsim}

\newcommand{\vect}[1]{\mbox{\boldmath${#1}$}}
\newcommand{\lmk}{\left(}
\newcommand{\rmk}{\right)}

\newcommand{\lkk}{\left[}
\newcommand{\rkk}{\right]}
\newcommand{\lla}{\left\langle}

\newcommand{\rra}{\right\rangle}

\newcommand{\ver}{{\vect r}}

\begin{document}
\title{LISA Measurement of Gravitational Wave Background Anisotropy: Hexadecapole Moment via a Correlation Analysis} 
\author{Naoki Seto and Asantha Cooray}
\affiliation{Theoretical Astrophysics, MC 130-33, California Institute of Technology, Pasadena,
CA 91125\\
E-mail: (seto,asante)@tapir.caltech.edu
}

\begin{abstract}
We discuss spatial fluctuations in the gravitational wave background arising from unresolved 
Galactic binary sources, such as close white dwarf binaries, due to the fact the galactic binary source
distribution is anisotropic. We introduce a correlation analysis of the
 two data streams of the Laser Interferometer Space Antenna (LISA) 
 to extract spherical harmonic coefficients, in an independent manner,
of the hexadecapole moment ($l=4$) related to the projected two-dimensional density distribution of
the binary source population. The proposed technique complements and improves over previous suggestions in the literature 
to measure  the gravitational wave background anisotropy based on the time modulation of data as LISA orbits around the Sun.
Such techniques, however, are restricted  only to certain combinations of spherical harmonic coefficients of the galaxy with no 
ability to separate them individually. With LISA, $m=2,3$ and $4$ coefficients of the
 hexadecapole ($l=4$) can be measured with signal-to-noise ratios 
at the level of 10 and above in a certain coordinate system. In addition to the hexadecapole coefficients, when combined with the time modulation analysis, 
the correlation study can also be used, in principle, to measure quadrupole coefficients of the binary distribution.
\end{abstract}
\maketitle

\section{Introduction}

In addition to gravitational waves from massive black holes at
cosmological distances,  
the Laser Interferometer Space Antenna (LISA; \cite{lisa}) is expected to
detect the galactic binary source background, such as close white dwarf binaries, with gravitational waves at frequencies between
0.1 mHz to 100 mHz \cite{webb}. At frequencies above 3 mHz, LISA will resolve a large number of galactic binaries \cite{Neletal01}, 
while at the low frequency end, the unresolved population will form a
confusion background of gravitational waves. This background is
expected to be anisotropic given the fact that galactic binaries trace the density distribution of our galaxy such that
the cumulative gravitational wave flux  is expected to be
 concentrated towards the highest density regimes such as the disk and the bulge of
the Milky Way. As LISA orbits around the Sun, its response will be
 sensitive to different regions on the sky and one expects
the data stream to modulate as a function of time. 
This modulation, in return, can be used to extract information related to the density
distribution and to reconstruct the anisotropy of the gravitational wave
background \cite{gia,Cornish:2001hg,Ungarelli:2001xu}. Note that the binary background can be considered as both a source of
noise or a signal; For example, by treating the background as a signal,
we can attempt to obtain certain information related to the Galactic structure through
anisotropies of the background \cite{gia,Cornish:2001hg,Ungarelli:2001xu}. On the other hand, if the background is
treated as a source of noise, we might also discuss how anisotropies affect the data analysis and
consider certain strategies to minimize any impact \cite{Seto04}.

Here, we treat the background as a signal and consider the presence of anisotropies and their potential measurement using a
correlation between the two data streams of LISA. For this purpose, following Ref. \cite{Seto04}, we introduce a
spherical harmonic moment analysis of LISA data in terms of spatial inhomogeneities in the two-dimensional 
projected density distribution when integrated over the observer to source distance. 
We also emphasize the subtle use of a freedom related to data combinations that also corresponds to an
effective rotation angle of LISA detectors in the detector plane. The technique we discuss here
compliments, and
improves, studies based on the time modulation of the data stream and allows one to extract individual coefficients of 
multipole moments related to the two-dimensional binary source distribution; 
Prior techniques were only restricted to combinations of coefficients related to the
monopole ($l=0$), quadrupole ($l=2$), and the hexadecapole ($l=4$). 
Considering LISA observational parameters, we show that 
certain coefficients of the hexadecapole multipole moments can be extracted with signal-to-noise ratios at the level of 10 and above. 

We also study the complimentary nature of the two methods involving time modulation and the correlation of data
streams and suggest that, in combination, one can also, in principle, extract information related to coefficients of the
quadrupole distribution. The only multipole moment coefficients that will remain individually undetermined
are ones of $l=0$,$m=0$ mode (monopole) and the $l=2, m=0$ mode related to the quadrupole. These two, however, can be determined,
in combination, as a sum. The correlation analysis would be one of the crucial 
elements for LISA's follow-on missions aiming to  detect weak
gravitational  wave background from the early universe
\cite{Ungarelli:2000jp,Allen:1996vm}. While it is expected that the  basic design of  any potential missions in the long term
is likely to be largely different from that of LISA, the correlation
analysis with  LISA data streams may present an important first step for any planned studies in the future that attempt to use similar procedures to both extract signals or to remove foreground noise.

The discussion is organized as follows: in the next Section, we outline the spherical harmonic formulation of the anisotropies
in the gravitational wave background and how the low order moments, $l=0,2$ and 4, are related to the LISA data streams.
In Section~3, we discuss the correlation between data streams as a way to extract information related to these anisotropies
beginning with a brief discussion of prior proposed techniques related to modulations in the data streams. In Section~4, We conclude
with a summary of our main results. For background information related to the technique presented in this paper, we refer the
reader to Ref.~\cite{Seto04}. 

\section{Formulation}

In this section, we first discuss the measurement of gravitational wave background with LISA detectors briefly 
and then discuss how this
data is related to the density distribution of the binary background, in terms of the spherical moments of
inhomogeneities in the projected distribution on the sky. For the purpose of this discussion, we make use of the
signal matrix involving the two data streams.

\subsection{Detector Response}
At the low frequency regime  of the LISA band,  two  separate modes, $(A,E)$, in gravitational waves can be
independently measured such  that the laser frequency noise is reduced to a level below other detector noises.
Furthermore, individual detector noises for these two modes do not correlate \cite{prince}.  We write these two data streams
$d_I~~(I=A,E)$, made with detector noise $n^D_{I}$, when the response to gravitational  waves, $h_{I}$, as
$d_I=h_I+n^D_{I}$. We define the detector noise spectrum  $S^D_{IJ}$ as $\lla n^{D*}_{I}(f) n^D_{J}(f^{'}) \rra=
\delta(f-f^{'})S^D_{IJ}(f)/2$.
The frequency dependence on noise is almost irrelevant for the present discussion and, hereafter, we omit the
explicit dependence for notational simplicity. Since the noise in two
modes are uncorrelated and have same magnitude due to the symmetric data combinations, we have the following
relation for the detector noise spectrum matrix
\beq
S^D_{AA}=S^D_{EE}, ~~~ S^D_{AE}=0.
\eeq

The responses  $h_{I}$  to  the low frequency gravitational  waves
for  two modes $(I=A, E)$ can be essentially
regarded as that of two  $90^\circ$-interferometers rotated by $45^\circ$
from each other, as shown in Fig.~1. To describe gravitational wave measurements, we take a reference coordinate
system $(X_D,Y_D,Z_D)$ from the detector system, where the $Z_D$-axis is normal to the detector plane.  
The gravitational waves behave as a Spin-2 field such that one can make linear combinations $[d_A(\phi_f), d_E(\phi_f)]$ 
from the original modes $[d_A,d_E]$ with a two dimensional  rotation matrix $R(\phi)$  as
\beq
\left( 
           \begin{array}{@{\,}c@{\,}}
           d_{A}(\phi_f)    \\
             d_{E}(\phi_f)   \\ 
           \end{array} \right)
 =R(2\phi_f)
\left( 
           \begin{array}{@{\,}c@{\,}}
           d_{A}    \\
             d_{E}   \\ 
           \end{array} \right). \label{trans}
\eeq
These new modes $[h_A(\phi_f), h_E(\phi_f)]$ are equivalent to  the responses
of two $90^\circ$-interferometers that are obtained by 
rotation of the original  modes  $(h_A,h_E)$ with an angle $\phi_f$ around the
$Z_D$-axis.  Based on this transformation,
one can easily confirm following relations for the detector noise spectrum  for an arbitrary rotational angle
\beq
S^D_{AA}(\phi_f)=S^D_{EE}(\phi_f)=S^D_{AA}=S^D_{EE},~~~S^D_{AE}(\phi_f)=0.
\eeq
As we discuss later, this freedom related to the effective rotation (or
data combination) is important for
the measurement of anisotropies related to the galactic gravitational wave background.

The signal matrix $S^B_{IJ}$ for the galactic binary background is defined as in the
case of detector  noise and is  given by the three-dimensional spatial distribution of gravitational-wave emitting
binaries, $\rho$, when integrated over the volume, as
\beq
S^B_{IJ}=P \int d^3\ver \rho (\ver) r^{-2}
F_{IJ}(\theta,\phi) \, ,
\label{sb}
\eeq
where $P$ is a normalization factor that is not important for our
analysis.   The weighing factor $r^{-2}$ in the integral is related to the fact that the gravitational wave
amplitude scales as $r^{-1}$, where $r$ is the radial distance from the detector.
Here, we assume that the frequency distribution and the spatial 
distribution of galactic gravitational wave emission rate  are
independent.  The functions $F_{IJ}$ represent the dependence of the
detector's response on the source direction, and are defined as 
\beq
F_{IJ}=\frac12\lkk \lmk \frac{1+\cos^2\theta }2 \rmk^2 p_1
+\lmk\cos\theta \rmk^2 p_2  \rkk.
\eeq
with $(p_1,p_2)=(\cos^2 2\phi, \sin^2 2\phi)$ for $F_{AA}$,
$(p_1,p_2)=(\sin^2 2\phi, \cos^2 2\phi)$ for $F_{EE}$, and
$(p_1,p_2)=(\sin 2\phi\cos 2\phi,-\sin 2\phi\cos 2\phi)$ for $F_{AE}$.
Using a spherical coordinate we can express  Eq.~(\ref{sb}) as \cite{gia}
\beq
S^B_{IJ}=\int d\Omega B(\theta,\phi)F_{IJ}(\theta,\phi),
\eeq
where the angular dependence of   the gravitational  wave background luminosity
$B(\theta,\phi)$ is defined through the radial integral as
\beq
B(\theta,\phi)\equiv P \int dr  \rho(r, \theta,\phi).\label{gwaniso}
\eeq 
Note the dependence on $r^{-2}$ factor in Eq.~\ref{sb} is removed due to the $r^2 dr$ factor involved in the $d^3\ver$ integral.
Thus, the projected two dimensional structure of the density distribution, as seen by gravitational waves, is simply
proportional to the column density of the binary population; this is also proportional to the
radial distance averaged density profile and, roughly speaking, the function
$B(\theta,\phi)$ can be regarded as the optical luminosity distribution on
the sky. 
As before, one can relate the signal matrix under a rotation such that $S^B_{IJ}(\phi_f)=R(2\phi_f)S^B_{IJ} R(-2\phi_f)$ for the
new data combination of $[d_A(\phi_f), d_E(\phi_f)]$. 

\subsection{Spherical Harmonic Expansion}

As shown in Fig.~1, the noise matrix $S^B_{IJ}$ is determined by the orientation of the
coordinate  system $K_D; (Z_D,Y_D,Z_D)$ which  changes with the rotation of LISA. 
Given this, we consider the spherical harmonic expansion under a fixed coordinate system 
$K_0 ;(X_0,Y_0,Z_0)$ since such a fixed system can aid in analyzing the gravitational wave
background \cite{Cornish:2001hg,Ungarelli:2001xu}. In this system, we set the $(X_0,Y_0)$-plane on the ecliptic and take $X_0$-axis in the direction of
the autumn equinox from the Sun. This is a reasonable choice considering
the LISA configuration \cite{lisa,Seto04}.   The two systems $K_D$ and $K_0$ are related by the Euler 
 angles $(\alpha,\beta,\gamma)$. Here $(\alpha,\beta)$ is the direction of
the $Z_D$-axis in the fixed $K_0$ system, while $\gamma$  is essentially 
degenerate with the freedom related to $\phi_f$-rotation  in
Eq.(\ref{trans}) around the $Z_D$-axis. Therefore, we use the angle
$\gamma$ hereafter as an independent variable that can be varied. 

The  angular dependence of the gravitational wave intensity $B(\theta,\phi)$  can be decomposed with  spherical harmonic
moments $Y_{lm}(\theta,\phi)$ (with the angular variables in the $K_0$
system) as   
\beq
B(\theta,\phi)=\sum_{l,m} B_{lm} Y_{lm}(\theta,\phi).
\eeq
Here, harmonic coefficients  $B_{lm}$ are given as
\beq
B_{lm}=\lla lm|B \rra,
\eeq
where we  have used the traditional notation for the inner product
$
\lla a | b\rra \equiv \int d\Omega a^*(\theta,\phi)
b(\theta,\phi) ,
$ 
and
an abbreviation $|lm \rangle\equiv |Y_{lm}(\theta,\phi)\rangle$.
Our goal in this subsection is to write down the matrix $S^B_{IJ}(\alpha,\beta,\gamma)$, in the
moving $K_D$ frame that is characterized by the Euler angles $(\alpha,\beta,\gamma)$, with coefficients $B_{lm}$
related to the background density distribution. 

First, the matrix $S^B_{IJ}(\alpha,\beta,\gamma)$  is formally expressed as 
\beq
S^B_{IJ}(\alpha,\beta,\gamma)=\lla F_{IJ}(\theta,\phi)|
U(\alpha,\beta,\gamma)^{-1}| B(\theta,\phi)\rra,
\eeq
where $U(\alpha,\beta,\gamma)$ is the rotation operator related to Euler
angles $\alpha,\beta$ and $\gamma$.
After some tedious but straight forward algebra (see, the Appendix A of Ref.~\cite{Seto04} for details), we write
\beqa
S^B_{AA}&=&\frac{2 {\sqrt \pi}}{5}B_{00}+\frac{4 \sqrt{\pi}}{7 \sqrt{5}}\sum_m D^2_{0m}
{\rm Re}[e^{im\alpha}B_{2m}]+\frac{\sqrt{\pi}}{105}\sum_m D^4_{0m} {\rm
Re}[e^{im\alpha}B_{4m}]  +\frac{2{\sqrt {\pi}} }{3 {\sqrt {70}}} \sum_m  D_{4m}^4 {\rm Re} [B_{4m} e^{(4\gamma+m\alpha)i}] \, , \nonumber \\
S^B_{EE}&=&\frac{2 {\sqrt \pi}}{5}B_{00}+\frac{4 \sqrt{\pi}}{7 \sqrt{5}}\sum_m D^2_{0m}
{\rm Re}[e^{im\alpha}B_{2m}]+\frac{\sqrt{\pi}}{105}\sum_m D^4_{0m} {\rm
Re}[e^{im\alpha}B_{4m}]  -\frac{2{\sqrt {\pi}} }{3 {\sqrt {70}}} \sum_m
D_{4m}^4 {\rm Re} [B_{4m} e^{(4\gamma+m\alpha)i}] \, , \nonumber \\
S^B_{AE}
&=&\frac{2 {\sqrt {\pi}}}{3{\sqrt {70}}} \sum_m  D_{4m}^4 {\rm Im} [B_{4m} e^{(4\gamma+m\alpha)i}]\, . \label{basic} 
\eeqa
As written, the two data streams from LISA detectors are only sensitive to the background $B_{lm}$ with  $l=0,2$ and 4
harmonics. This is simply due to the angular dependence of the $F_{IJ}(\theta,\phi)$ term such that $\lla F_{IJ}| lm \rra$ is
only non-zero when $l=0,2$ and 4. Unlike electromagnetic radiation observations, such as say that of photons in the
cosmic microwave background, where, in principle, all multipole moments of the inhomogeneous radiation field is
measurable,  with gravitational waves, one is only restricted to these
low order  
moments of the underlying field due to the poor directional
characterization of the gravitational wave detectors. One can potentially extract additional multipole moments of the gravitational wave background with
more data combinations involving, say, an additional configuration such
as another set of 3 spacecrafts \cite{Cornish:2001hg,Ungarelli:2001xu}.
Since the LISA configuration is fixed to three spacecrafts, we do not discuss such possibilities further.
Note that our expressions in Eq.(\ref{basic}) are valid at  the long wavelength limit of LISA
relevant for the Galactic confusion background. At higher
frequency regime, we expect the dependence of these terms in terms of
harmonic coefficients to change  \cite{Cornish:2001hg,Ungarelli:2001xu}.

The off-diagonal  element of the noise matrix $S^B_{AE}$ is non-zero due to
the $l=4$ coefficients of the binary background. The above equations related to $S^B_{IJ}(\alpha,\beta,\gamma)$
are essentially same as eqs.(A13)-(A15) in Ref.~\cite{Seto04}. Here, in comparison,
we have rewritten the spin-weight harmonics in that paper 
with matrix elements $D^l_{sm}(\beta)=\lla ls|U(0,\beta,0)^{-1}| lm \rra$ with
\beq
 {}_{-s}Y_{lm} (\beta,\alpha) \sqrt{\frac{4\pi}{2l+1}}
=D^l_{sm}(\beta) e^{i m \alpha} \; 
\eeq
and elements  $D^l_{sm}(\beta)$ given by the Jacobi polynomials
$P^{(s+m,s-m)}_{l-s}$.  When $s \ge m$, their explicit forms are \cite{Allen:1996gp}
\beq
D^l_{sm}(\beta)=(-1)^{l-s} \lkk\frac{(l-s)!(l+s)!}{(l-m)!(l+m)!}
\rkk^{1/2} \lmk \cos \frac{\beta}2  \rmk^{s+m} \lmk \sin \frac{\beta}2  \rmk^{s-m} P^{(s+m,s-m)}_{l-s}(-\cos\beta) \, ,
\eeq
while for $s<m$, $D^l_{sm}(\beta)=(-1)^{s-m}D^l_{ms}(\beta)$ 

In the case of observations with LISA, the angle $\beta$ is fixed to the value of $\pi/3$. This leads to
the restriction that,  in principle,  one cannot separate information related to $B_{00}$ and $B_{20}$. 
The angle $\alpha$ is given as $\alpha=2\pi t$ with orbital time $t$ in units of year (measured from the Autumn equinox).
For the Galactic binary distribution $\rho(\ver)$ we use the triaxial
model given in \cite{Binney:1996sv} and used in \cite{Seto04}. 
The relevant coefficients $B_{lm}$ and $D^l_{sm}$ are presented in
Tables~I and II. 

While we primarily discuss the background from our galaxy, it is worth considering the possibility that anisotropies
can exist due to gravitational wave emission from binaries in other galaxies of the Local group, such
as the Andromeda galaxy (M31) or the Large Magellanic Cloud (LMC). Their contribution to  the anisotropy $B(\theta,\phi)$ is roughly 
proportional to $\rm (mass)/(distance)^2$. The characteristic distance
is 10kpc for the Milky Way galaxy, 50kpc for LMC and 700kpc for M31. As
for the mass, M31 is comparable to Milky Way, but LMC is $\sim 50^{-1}$
times smaller. Given these parameters, we find that the  contribution to anisotropies of the gravitational wave
background from other member of the local group is negligible.  This is
also apparent from the fact that the optical luminosity of these and other
 galaxies are much lower than the Milky Way.

\section{Extraction of Anisotropy Parameters}

Having established the relation between components of the LISA data stream matrix and coefficients of the
binary background anisotropy, we now discuss the individual measurement of these anisotropy coefficients based on
various combinations of terms in the data matrix. First, we briefly outline the previous method based on the
time modulation of the data stream and consider the relation between anisotropy coefficients there and the
ones we discuss here.

\begin{figure}
  \begin{center}
\epsfxsize=7.cm
\begin{minipage}{\epsfxsize} \epsffile{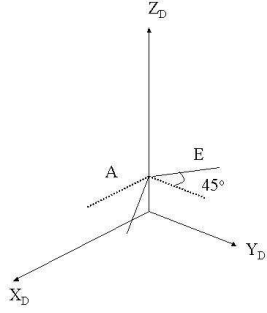} \end{minipage}
 \end{center}
  \caption{ Definition of the detector coordinate $(X_D,Y_D,Z_D)$. The
 $Z_D$-axis is normal  to the detector plane. A and E modes can be
 regarded as two L-shaped detectors rotated by $45^\circ$.}
\end{figure}

\begin{figure}
  \begin{center}
\epsfxsize=8.cm
\begin{minipage}{\epsfxsize} \epsffile{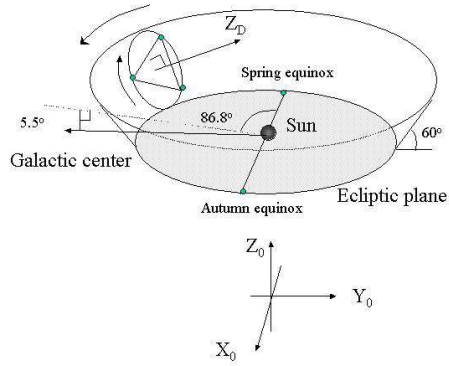} \end{minipage}
 \end{center}
  \caption{Configuration of LISA and the definition of the fixed ecliptic
 coordinate $(X_0, Y_0, Z_0)$. The $Z_0$-axis is normal to the ecliptic
 plane and  $X_0$-axis is oriented to   the autumn
 equinox from the Sun. } 
\end{figure}

\subsection{Time Modulation of LISA Data}

Previous studies on the measurement of anisotropies $B_{lm}$ are
based on the time modulation of the spectrum $S^B_{AA}$
\cite{gia,Cornish:2001hg,Ungarelli:2001xu} (see also
\cite{Allen:1996gp}). These techniques do not use
(i) other modes $S^B_{AE}$ or $S^B_{EE}$ and (ii) the freedom to adjust one of the Euler angles,
$\gamma$, a priori. In time modulation studies, the angle $\gamma$ simply follows the motion of
LISA and given as $\gamma=-2\pi t+\gamma_0$ where $\gamma_0$ is some constant
angle (here, we put $\gamma_0=0$ for simplicity).  We briefly follow the underlying idea behind time-modulation method 
and highlight certain problems for the signal  extraction related to $B_{lm}$.
 The function  $S^B_{AA}$ can be expanded as
$S^B_{AA}(t)=S^B_{AA}(\alpha=2\pi t, \beta=\pi/3, \gamma=-2\pi
t)=\sum_{n=-8}^8C_n e^{2\pi i n
t}$. Since $C_n^*=C_{-n}$, as one is observing a real realization on the sky, 
we do not discuss $C_n$ with $n<0$.  The explicit forms of the
coefficients $C_n$ are easily derived from Eqs.(\ref{basic})
\beq
C_0=\frac{2 {\sqrt \pi}}{5}B_{00}+\frac47\sqrt{\frac{\pi}5}D^2_{00} B_{20}+{\frac{\sqrt\pi}{105}}
D^4_{40}B_{00}+\frac23{\sqrt{\frac{\pi}{70}}}D^4_{44} {\rm Re}[B_{44}],
\eeq

\beq
C_1=\frac47\sqrt{\frac{\pi}5}D^2_{01} B_{21}+\frac{\sqrt \pi}{105}
D^4_{01}B_{41}-\frac13{\sqrt{\frac{\pi}{70}}}D^4_{43}B_{4-3},
\eeq

\beq
C_2=\frac47\sqrt{\frac{\pi}5}D^2_{00} B_{22}+\frac{\sqrt \pi}{105}
D^4_{02}B_{42}+\frac13{\sqrt{\frac{\pi}{70}}}D^4_{42}B_{4-2},
\eeq

\beq
C_3=\frac{\sqrt \pi}{105}
D^4_{03}B_{43}-\frac13{\sqrt{\frac{\pi}{70}}}D^4_{41}B_{4-1}, ~~~
C_4=\frac{\sqrt \pi}{105}
D^4_{04}B_{44}+\frac13{\sqrt{\frac{\pi}{70}}}D^4_{40}B_{40},
\eeq

\beq
C_5=
-\frac13{\sqrt{\frac{\pi}{70}}}D^4_{4-1}B_{41},~~~
C_6=
\frac13{\sqrt{\frac{\pi}{70}}}D^4_{4-2}B_{42},~~~
C_7=
-\frac13{\sqrt{\frac{\pi}{70}}}D^4_{4-3}B_{43},~~~
C_8=
\frac13{\sqrt{\frac{\pi}{70}}}D^4_{4-4}B_{44}.
\eeq

Assuming that the detector noise $S^D_{AA}$ is significantly smaller  than the
background  $S^B_{AA}$ in a  frequency
bandwidth  $\Delta f$,  the signal-to-noise ratio for each coefficients
 $(n\ne 0)$ is given as \cite{gia,Ungarelli:2001xu}
\beq
\lmk \frac{S}{N} \rmk_n=\sqrt2 \frac{|C_n|}{C_0}\sqrt{\Delta f
T}.\label{cnsn}
\eeq
where $T$ is the observational period.
Statistics for the
measurement of the time dependent signal $C_n$ $(n\ne 0)$ would be 
largely different from that  of  the constant part $C_0$.  The coefficients $C_n$ are
complex numbers but their phases depend on the choice of the orientation of the
$X_0$-axis that is somewhat arbitrary in the ecliptic plane. 
Therefore, we summed up the signal  to noise ratios of their real and
imaginary parts in Eq.~(\ref{cnsn}).  
The numerical prefactor of $\sqrt{2}$ comes from the fact that we assume the noise contributions to the
real and imaginary parts of these coefficients are related to the total noise, $N_T$ such that
$N_R = N_I = N_T/\sqrt{2}$ without correlation between them. 
For our model parameters related to the binary distribution, we
have  
\beq
\lmk \frac{S}{N} \rmk_1=50,
\lmk \frac{S}{N} \rmk_2=73,
\lmk \frac{S}{N} \rmk_3=3.6,
\lmk \frac{S}{N} \rmk_4=2.2,
\lmk \frac{S}{N} \rmk_5=0.69,
\lmk \frac{S}{N} \rmk_6=0.76,
\lmk \frac{S}{N} \rmk_7=0.23,
\lmk \frac{S}{N} \rmk_8=0.084,
\eeq
with $T=10^8$sec and $\Delta f=10^{-3}$Hz.  Thus, one 
can obtain only $C_1$ and $C_2$ with sufficient statistical
significance. The above signal-to-noise ratios are  consistent with previous
analyses when considering the difference between models for the Galaxy density distribution between these
calculations and the present one \cite{gia}.  Since $C_1$ and $C_2$ depend on
certain combinations of $B_{lm}$, one cannot separate out the multipole moment coefficients related to the gravitational
wave background by simply measuring the time modulation components in the Fourier space.
Moreover, in terms where certain $B_{lm}$ 
coefficients are simply proportional to $C_n$ coefficients, especially when
$n > 5$, the dependence is significantly reduced through small coupling
coefficients.
For example, the information related to $B_{4\pm3}$  appears  in $C_7$ with the factor
$-D^4_{4-3}\sqrt{\pi/70}/3 =-0.0044$. Once can consider combinations of $C_n$ to extract a certain
$B_{lm}$, but, again for example for the case with $B_{4\pm3}$, the dependence   in $C_3$ is through 
$ D^4_{03}\sqrt{\pi}/105=-0.0081$ and in $C_1$ through
$-D^4_{43}\sqrt{\pi/70}/3 =-0.036$.  The former factor (in $C_7$)
is smaller  than the latter (note $D^4_{4-3}\ll  D^4_{40}, D^4_{43}$; Table II),
but information related to $B_{43}$ in the latter remains to be
 degenerate with  other $B_{lm}$ coefficients such that one cannot easily find the necessary combinations.
The situation is also similar for $B_{4\pm2}$ or $B_{4\pm4}$.
The factor $D^4_{4-4}\sqrt{\pi/70}/3$ in $C_8$ is extremely small, and
it is impossible to  determine $B_{44}$ form the coefficient $C_8$.

 From these examples, we learn two things: We should try to
improve the signal-to-noise ratio for  $B_{lm}$ coefficients by taking a suitable
combination with respect to the matrix element $D^4_{lm}$, and handle the degeneracy of $B_{lm}$ coefficients with the use of
other modes. 

In the next subsection, we study a method for signal analysis where we pay
special attention to these two points. 

\subsection{Correlation Analysis}

We use the correlation analysis $\int S_{AE}(t)dt$ to  measure the
parameter $B_{4m}$.  By setting $\gamma=-\pi/2 m' t+\pi/8$ 
$(m' \ge 0)$, for each $m'$ from 0 to 4, a priori --- since we have this freedom to specify this angle before hand ---
one obtains, under the scenario that $\alpha=2\pi \; t$ and after integrating the signal $S_{AE}(t)$ over an observational
period $T$ (an integer in units of one year),
\beqa
\lla S^B_{AE} \rra^{m'}_R &\equiv& \int S^B_{AE}(t)\; dt \nonumber \\
&=&\int dt\; \frac{2 {\sqrt {\pi}}}{3{\sqrt {70}}} \sum_m  D_{4m}^4 {\rm Im} [B_{4m} e^{i(2\pi(m-m')t+\pi/2)}] \nonumber \\
&=& -\frac{2 {\sqrt {\pi}}}{3{\sqrt {70}}} D_{4m'}^4 {\rm Re}[B_{4m'}] \, ,
\eeqa
while the imaginary component of the same quantity can be obtained by setting $\gamma=-\pi/2 m' t$ such that
$\lla S^B_{AE} \rra^m_I \propto {\rm Im}[B_{4m}]$. Thus, the substitution for $\gamma$ has the advantage that 
one can now individually get real  and imaginary parts of $B_{4m}$, while at the same time, also making use of the
substantially large  coupling coefficients of $D^4_{4m}$ $(m\ge 0)$ that are associated with harmonic coefficients.

When we  neglect the detector noise and the frequency dependence of
anisotropies, which would be tolerable for the present study,  the signal-to-noise ratio of the correlation analysis, for
an individual $l=4$ and $m$ coefficient, is
\beq
({\rm S/N})_{4m}= \lmk \int dt \int df \frac{[\lla S^B_{AE} \rra^m]^2}{S_{AA}(t)S_{EE}(t)}   \rmk^{1/2} \, ,
\eeq
where we have now combined the real and imaginary parts $[\lla S^B_{AE} \rra^m]^2=[\lla S^B_{AE} \rra_R^m]^2+[\lla S^B_{AE} \rra_I^m]^2$.
We have numerically calculated these signal-to-noise ratios for each $m$ value using the same density distribution
model for the galactic binaries as before and we obtain following values for
$(S/N)_{4m}$ for the measurement of $B_{4m}$ coefficients, separately, as
\beq
(S/N)_{40}=1.1,~~
(S/N)_{41}=3.7,~~
(S/N)_{42}=12,~~
(S/N)_{43}=13,~~
(S/N)_{44}=12,~~
\eeq
when $\Delta f=10^{-3}$Hz and $T=10^8$sec.
Thus, as estimated, we expect one can  estimate the coefficients
 $B_{42}$, $B_{43}$ and $B_{44}$ accurately, while  upper
limits can be established for $B_{40}$ and $B_{41}$.
Then, using the previous time modulation method,  one can further  estimate $B_{21}$ form $C_1$ coefficient 
and $B_{22}$ from $C_2$ coefficient.  Since coefficients $C_1$ and $C_2$  are dominated by $B_{12}$ and $B_{22}$ respectively,
this determination can be carried out with sufficient signal-to-noise ratio.

So far, we have studied the correlation analysis for estimating $B_{4m}$ individually for all $m$ values and
the combination of time modulation related to $S^B_{AA}$ and the correlation analysis to extract certain the coefficients
related to the quadrupole. One can also use the time modulation method alone, but with combinations of $S^B_{AA}$, 
$S^B_{EE}$, and $S^B_{AE}$. An example is  the modulation of $S^B_{AA}-S^B_{EE}$. Again,
the freedom related to the angle $\gamma$ is important. For example by setting $\gamma=2\pi t (-m/4+2)$ or $2\pi t
(-m/4+2)+\pi/8$, we can estimate the real and imaginary parts of $B_{4m}$ 
from the coefficient $C_8$ related to the combination $S^B_{AA}-S^B_{EE}$. By making numerical estimates, we find that the
signal-to-noise ratio for such a measurement is similar to the one estimated above with the correlation analysis.

While the combination with time modulation for $S_{AA}^B$ alone do lead to certain coefficients of $B_{4m}$, we find that
the correlation analysis improves the correlation method by providing a mechanism to extract all
coefficients. Moreover, the correlation method compliments techniques related to the time modulation and we find that,
in combination, the methods can be used to extract most coefficients related to the $l=2$ and $4$ moments of the
galactic distribution of binaries. The only coefficients that are not independently measured is $B_{00}$ and
$B_{20}$, though the two can be measured in combination as a sum. As
mentioned earlier, signal analysis for the $C_0$ mode is different from other  $C_n$ modes with $n \ne 0$. For
this purpose, data streams other than $A$ and $E$ modes, {\it e.g.} Sagnac, would be important to calibrate
the detector noise \cite{Hogan:2001jn}. 

\begin{table}[t]
\begin{center}
\begin{tabular}{|c||c||c|c|c||c|c|c|c|c||c|c|c|c|c|c|} \hline
$l,m$ & $0,0$  & 2,0 & 2,1 & 2,2 & 
 4,0 & 4,1 & 4,2 & 4,3 & 4,4 
\\ \hline
$B_{lm}/B_{00}$ &1   &-0.243 & -0.292+0.077$i$
 &  -0.647-0.038$i$& 0.052 & -0.076+0.076$i$ &0.289-0.012$i$  & -0.329-0.073$i$ &
 0.504+0.061$i$  
\\ \hline
\end{tabular}
\caption{ Spherical harmonic coefficients of  the Galactic binary
 background. Note that  $B_{l-m}=(-1)^m B_{lm}^*$. 
}
\end{center}
\end{table}

\begin{table}[t]
\begin{center}
\begin{tabular}{|c|c|c|c|c|} \hline
   $D^2_{0,-2}$  & $D^2_{0,-1}$   & $D^2_{0,0}$ & $D^2_{0,1}$ &
 $D^2_{0,2}$ \\ \hline
 0.459 & 0.530 & -0.125 & -0.530 & 0.459
\\ \hline
\end{tabular}

\begin{tabular}{|c|c|c|c|c|c|c|c|c|} \hline
   $D^4_{0,-4}$  & $D^4_{0,-3}$  & $D^4_{0,-2}$ & $D^4_{0,-1}$  & $D^4_{0,0}$ & $D^4_{0,1}$ &
 $D^4_{0,2}$ & $D^4_{0,3}$ &
 $D^4_{0,4}$ \\ \hline
 0.294 & 0.480 & 0.222 & -0.303 & -0.289  & 0.303 & 0.222 & -0.480 & 0.294
\\ \hline
\end{tabular}

\begin{tabular}{|c|c|c|c|c|c|c|c|c|} \hline
   $D^4_{4,-4}$  & $D^4_{4,-3}$  & $D^4_{4,-2}$ & $D^4_{4,-1}$  & $D^4_{4,0}$ & $D^4_{4,1}$ &
 $D^4_{4,2}$ & $D^4_{4,3}$ &
 $D^4_{4,4}$ \\ \hline
 0.004 & 0.019 & 0.062 & 0.152 & 0.294  & 0.456 & 0.558 & 0.516 & 0.316
\\ \hline
\end{tabular}
\caption{Matrix elements $D^l_{sm}(\beta)$ for $\beta=\pi/3$.
}
\end{center}
\end{table}

\section{Summary}

In this paper, we have discussed spatial fluctuations in the gravitational wave background arising from unresolved 
Galactic binary sources, such as close white dwarf binaries, due to the fact the this galactic source
distribution, when projected on the sky, is anisotropic. We introduce a spherical harmonic analysis of this anisotropy  based on
a correlation study of the two  data streams of LISA and propose the
measurement of individual spherical harmonic coefficients related to the hexadecapole moment ($l=4$), in addition to
coefficients of the quadrupole ($l=2$). 

The proposed technique complements and improves over previous suggestions in the literature to 
measure  the gravitational wave background anisotropy based on the time
modulation of the single data stream ({\it e.g.} $S^B_{AA}(t)$) as LISA orbits around the Sun
and restricted  only to combinations of spherical harmonic coefficients with no ability to separate them.
The correlation analysis, by making use of a single freedom related to the rotation angle of detectors in the
detector plane or similarly how data are combined,  provides a method to independently extract multipole
coefficients of the hexadecapole moment of the background. The information related to the quadrupole comes from
the correlation analysis in combination with the time modulation. The only independently undetermined coefficients
turn out to be that of the monopole and the $m=0$ mode of the quadrupole, though the two can be determined as a sum.

With LISA, we have shown that certain coefficients of the hexadecapole can be measured with signal-to-noise ratios at the level of 10
and above. While LISA will present a first step towards a correlation analysis of the gravitational wave background,
future missions are likely to exploit this in detail both as a source of
signal, to extract spatial information, 
and noise, to decrease noise to detect almost isotropic backgrounds such
as the one related to primordial gravitational waves. This would certainly be the case at low
frequencies as the foreground is likely to be the dominant source of anisotropy. At few mHz frequencies,
 the fluctuation amplitude related to the extragalactic component
is expected to be at the level of a few percent, but restricted to arcminute angular scales
where galaxy density field is observed to vary. At $\mu$Hz frequencies, where the merging massive black holes are
likely to be the domiant foreground source, the anisotropy will be dominated by the shot-noise related to the finite number 
density of such binaries on the sky.
On the other hand, the anisotropy of the primordial gravitational wave background is expected to have a pattern close to that
of the cosmic microwave background, with a fluctuation level of
$\delta \sim 10^{-5}$, at large angular scales.  
A confirmation of their similarity in anisotropy can be used as a
proof on the primordial nature of the background. For a direct detection of spatial fluctuations in the
gravitational wave background, in  the
optimistic case where the detector noise is dominated by the monopole
mode of the background, one is forced to gravitational wave frequencies
at the level of $f\sim \Delta f\sim 10^4 ({\rm SN}/10)^2 (\delta/10^{-5})^{-2}(T/3{\rm yr})$Hz.  
Thus, for a detection of the anisotropy at a level $\delta\sim 10^{-5}$ over a realistic observational period with significant signal to  noise ratio, the relevant frequencies are at $f\gsim
10$kHz. Interestingly, at this high end of the gravitational wave spectrum, 
there are no viable astrophysical sources and suggests that, in future, one may in fact be able to
extract anisotropy associated with the primordial gravitational wave background.

\acknowledgments
This work was supported in part by DoE DE-FG03-92-ER40701,
a senior research fellowship from the Sherman Fairchild foundation (AC),
NASA grant NAG5-10707, and the JSPS fellowship to research abroad (NS).

\end{document}